\documentstyle[prd,aps,preprint]{revtex}
\begin{document}
\draft

%
%  Uncomment following two lines and one below for 2 column format.
%@twocolumnfalse\endcsname

\preprint{Nisho-99/1} \title{Axionic Boson Stars\\
 in Magnetized Conducting Media} 
\author{Aiichi Iwazaki}
\address{Department of Physics, Nishogakusha University, Shonan Ohi Chiba
  277,\ Japan.} \date{January 8, 1999} \maketitle
\begin{abstract}
Axions are possible candidates of dark matter 
in the present Universe. They have been argued to form
axionic boson stars with small masses 
$10^{-14}M_{\odot}\sim 10^{-11}M_{\odot}$.
Since they possess oscillating electric fields 
in a magnetic field,
they dissipate their energies in magnetized
conducting media such as white dwarfs or neutron stars. 
At the same time the oscillating electric fields generate 
a monochromatic radiation with energy equal to mass of the axion.
We argue that 
the effect of the energy dissipation can be seen in the old white dwarfs. 
In particular,
We show that colliding with
sufficiently cooled white dwarfs, plausible candidates of MACHO, 
the axionic boson stars dissipate their energies in the dwarfs 
and heat up the dwarfs. Consequently the white dwarfs in the halo 
can emit detectable amount of thermal radiations with the collision.
On the other hand, the monochromatic radiations can be seen only during the 
collision; a period of the dwarf passing the axionic boson star.
Assuming that MACHO are dark white dwarfs,
we show that there is a threshold in luminosity function of the white 
dwarfs below which 
the number of the white dwarfs in the halo increase discontinuously.
The threshold in the luminosity function is 
expected to be located around $10^{-5.5}L_{\odot}\sim 10^{-7}L_{\odot}$.
Its precise value is determined by the mass of the axionic boson 
 stars dominant in the halo.

%A characteristic feature in the observation of these radiations is that 
%we first observe the monochromatic radiations in a short period
%and then we observe 
%the thermal radiations which are emitted until the dissipation 
%energy is exhausted.  
%Using a recent evaluation of the population of the white dwarfs
%as candidates of MACHO, we estimate     
%the event rate of the collisions and find that
%the rate is large for the collision to be detectable.
\end{abstract}
%\pacs{73.61.-r,73.20.Dx,73.40.Hm,73.40.Gk}
%\pacs{14.80.Mz, 98.80.Cq, 95.35.+d, 05.30.Jp, 97.20.Rp, 98.70.-f}   
%Boson Star 
%\hspace*{3cm}}
\vskip2pc
%%%%%%%%%%%%%%%%%%%%%%%%%%%%%%%%%%%%%%%%%%%%%%%%
\section{Introduction}
The axion is the Goldstone boson associated with 
Peccei-Quinn symmetry\cite{PQ}, 
which was introduced to solve naturally the strong CP problem. 
In the early Universe some of the axions
condense and form topological objects\cite{kim,text}, i.e. 
strings and domain walls, 
although they decay below the temperature of QCD phase transition. 
After their decay, however, they have been 
shown to leave a magnetic field\cite{iwa} 
as well as cold axion gas as relics in 
the present Universe; the field is a candidate of a primordial
magnetic field supposed to lead to galactic magnetic fields observed 
in the present Universe.

In addition to these topological objects,  
the existence of axionic boson stars has been argued\cite{hogan,kolb}.
It have been shown numerically\cite{kolb} that in the early Universe, 
axion clumps are formed around the period of 
$1$ GeV owing to both the nonlinearity of an axion potential leading to 
an attractive force among the axions and 
the inhomogeneity of coherent axion oscillations. 
Namely, when the temperature of the Universe decreases   
and the axion potential is generated by QCD instantons, 
the inhomogeneity of the coherent axion oscillation 
on the scale beyond the horizon gives rise to 
localized clumps due to the attractive force of the potential. 
These clumps are called axitons 
since they are similar to solitons in 
a sense that its energy is localized. Then,
the axitons contract gravitationally to axionic boson stars\cite{kolb2,real}
after separating out from the cosmological expansion.
They are solitons of coherent axions bounded gravitationally.
( The axions are represented with a real scalar field, which 
has been shown to possess solutions of oscillating boson stars\cite{real}. )
The masses of the axionic boson stars have been estimated 
roughly to be order of 
$\sim 10^{-12}M_{\odot}$. 
Eventually we expect that in the present Universe, 
there exist the axionic boson stars
as well as the axion gas as dark matter candidates. 
It has been estimated\cite{fem} that a fairly amount of the fraction of 
the axion dark matter is composed of the axionic 
boson stars.

In this paper we wish to point out an intriguing 
observable effect associated with 
the coherent axionic boson stars; we call them axion stars. 
Namely, they 
dissipate their energies in magnetized conducting media 
such as magnetic white dwarfs or neutron stars so that
the temperature of the media increases and thermal radiations are emitted. 
The phenomena are caused by oscillating electric fields generated by the  
axion field
under external magnetic fields; 
their frequency is given by the mass of the axion. 
The electric fields induce electric currents
in the conducting media and loose their energies owing to the existence of 
electric resistances. 
Consequently the axion stars dissipate their energies in the 
magnetized conducting media.
Although the electric fields themselves are small,
the total amount of the energy dissipation is very large 
because the dissipation
arises all over the volume of the axion stars or the volume of 
the magnetized conducting media: 
Radii of the white dwarfs ( the neutron stars ) are typically  
$10^9$cm ( $10^6$cm ), while 
radii of the axion stars of 
our concern are such as $10^6\mbox{cm}\sim10^{10}\mbox{cm}$.
Consequently detectable amount of radiations are expected 
from the media heated in this way.
In particular we are concerned with old white dwarfs which are plausible 
candidates of MACHO\cite{macho} and have been cooled sufficiently; 
they are dark enough 
to be invisible. 
Thus the effect of the 
energy dissipation of the axion star in such white dwarfs possessing 
small specific heats 
is so large
that the luminosity of the dwarfs after 
the collisions with the axion stars increases so as for the dwarfs to be 
observable.

In particular, we point out that there is a threshold in luminosity 
function of the halo white dwarfs ( luminosity function 
describes, roughly speaking, the number density of the white dwarfs 
as a function of luminosity ). Namely 
the luminosity function increases discontinuously below the threshold 
of a certain luminosity. To explain it, we first point out that 
the white dwarfs in the halo are sufficiently old so that their luminosity 
is quite low to be invisible. Especially, according to a recent cooling model,
the white dwarfs with helium
rich atmosphere are expected to have been cooled so that their luminosity 
is less than $10^{-7}L_{\odot}$ at their age $\sim 1.2 \times 10^{10}$ years.
Here we assume that MACHO are such white dwarfs with helium rich atmosphere.
Thus they are invisible with present observational apparatus unless 
they are located quite near the sun.   
Then since they have few internal thermal energies,  
their temperatures increase dramatically with the dissipation of 
the energy of the axion star in the white dwarfs. 
We show that the luminosity gained 
by the white dwarfs with the collision is about 
$10^{-5.5}L_{\odot}\sim 10^{-7}L_{\odot}$ depending on 
the mass of the axion star; the mass of our concern is 
$10^{-11}M_{\odot}\sim 10^{-14}M_{\odot}$ ( $L_{\odot}$ denotes
the luminosity of the sun ).
Therefore, this luminosity is expected to be 
the threshold luminosity of the white dwarf 
luminosity function in the halo.

In addition to these thermal radiations from the magnetized medium, 
the oscillating electric fields associated with 
the axion star in the magnetized medium 
generate monochromatic radiations with 
energy equal to the mass of the axion. Since the electric fields arise only 
during the axion star being exposed to an external magnetic field of the 
magnetized media, the radiations are emitted only in a period of 
the media passing the axion star. Therefore, we expect that we first observe 
the monochromatic radiations in a short period and then observe the thermal 
radiations which are emitted until the dissipation energy deposited by the 
axion star is exhausted. This thermal radiations are observed possibly
as a nova 
which is an old white dwarf heated with the collision.
Here we should mention that since the rate of the collision between the white 
dwarfs and the axion stars is much small, it seems difficult that we observe 
the monochromatic radiations.

Because the strength 
of the electric fields is proportional to the strength of the magnetic field,
the phenomena are distinctive
of the strongly magnetized media.
We show that the amount of the energy dissipated in white dwarfs, 
for instance, with mass $\sim 0.5\,M_{\odot}$ and with 
magnetic field larger than $10^{5}$ Gauss is approximately given by 
$10^{42}$ erg $M_a/10^{-12}M_{\odot}$ 
where $M_a$ ( $M_{\odot}$ ) is the 
mass of the axion star ( the sun ).    
Namely in almost of all cases, whole energy of the axion star is dissipated
in such white dwarfs.
In our discussions we neglect, for simplicity, 
gravitational effects of the magnetized
conducting media
on the axion stars when they collide with 
the media. Later we 
discuss on this point.

In section (2) we present our solutions of the axionic boson stars
with small masses. The axion field of the solutions 
oscillates with the frequency of the axion mass. 
We show explicitly relations between the mass and the 
radius of the axion star, which are exploited for the estimation of 
the amount of the energy dissipation in magnetized conducting media.
In section (3) we describe an intriguing phenomenon 
that the axionic boson star induces 
an oscillating electric field when it is exposed to an external magnetic field.
This electric field induces electric currents carried 
by ordinary charged particles in the 
magnetized conducting media. Thus the axion stars 
dissipate their energies in such media. 
The axion star in the magnetic field possesses  
another oscillating electric current even in 
nonconducting media such as vacuum
owing to an interaction of the axion field with the electromagnetic fields.
The current\cite{Si} is composed of the axion field 
although the axions are neutral.
Owing to these two types of the oscillating currents, 
the axion stars dissipate their energies
in the conducting media. Thus the media are heated so that thermal 
radiations are emitted. The application of these phenomena is discussed 
in section (4) and section (5) where white dwarfs and 
neutron stars are discussed as magnetized conducting media, respectively.
separately.  
We summarize our results in the final section (6).

\section{Axionic Boson Star}
Let us first explain our solutions of the axionic boson stars.
Originally Seidel and Suen \cite{real} 
have found solutions of a real scalar axion field, $a$,
coupled with gravity. Their solutions represent 
spherical oscillating axion stars
with masses of the order of $10^{-5}M_{\odot}$; the solution   
possesses oscillation modes with various frequencies.
( Static regular solutions have been shown 
not to exist in the massless real scalar field coupled with gravity. 
Even in the massive 
real scalar field such solutions have not yet been obtained. )
Although these axion stars are oscillating, they are stable solitons
composed of the axions coupled with gravity; they are similar to 
the ``breather'' solution of the ( 1 + 1 )-dimensional sine-Gordon model. 
Axion stars of our concern, on the other hand, are ones with much smaller 
masses, $\sim 10^{-14}M_{\odot}$. 
This is because according to arguments of 
Kolb and Tkachev\cite{kolb,kolb2,fem}
the axion stars produced in early Universe 
have masses typically such as $10^{-12}M_{\odot}\Omega_ah^2$ where
$\Omega_a$ is the ratio of the 
axion energy density to the critical density in the Universe and 
$h$ is Hubble constant in the unit of $100$ km s$^{-1}$ Mpc$^{-1}$.
They have been produced after the period of QCD phase transition
with axions contracting due to both 
effects of the gravitational attraction and the attraction
of the axion potential. 
  
So, in order to find such solutions
and to obtain explicit relations 
among the parameters, e.g. radius, $R_a$, mass, $M_a$, e.t.c., 
of these axion stars,  
we have numerically obtained solutions 
of the spherical axionic boson stars \cite{iwaza,real,re} in a 
limit of a weak gravitational field. Relevant equations are 
a free field equation
of the axion and Einstein equations,

\begin{eqnarray}
\label{a}
\ddot{a}&=&\frac{(\dot{h_t}-\dot{h_r})\dot{a}}{2}+a''
+(\frac{2}{r}+\frac{h_t'-h_r'}{2})a'-m_a^2a\quad,\\
\label{h_t}
h_t'&=&\frac{h_r}{r}+4\pi Gr(a'^2-m_a^2a^2+\dot{a}^2)\quad,\\
\label{h_r}
h_r'&=&-\frac{h_r}{r}+4\pi Gr(a'^2+m_a^2a^2+\dot{a}^2)\quad,
\end{eqnarray}
where we have assumed gravity being small, i.e. $h_{t,r}\ll 1$ 
so that the metric is such that 
$ds^2=(1+h_t)dt^2-(1+h_r)dr^2-r^2(d\theta^2+\sin^2\theta d\phi^2)$;
$r,\theta$, and $\phi$ denote the polar coordinates.
The first equation is the equation of the axion field $a$.
The second and the third equations are Einstein equations.  
A dot ( dash ) indicates a derivative in time $t$ ( $r$ ). $G$ ( $m_a$ ) 
is the graviational constant ( the mass of the axion ).
The potential term $\sim \sin(a)$ of the axion has been neglected
because an amplitude of the field $a$ is sufficiently small for 
nonlinearity not to arise since the mass of the axion 
star is small enough. Actually the masses we are concerned with are 
such as $\sim 10^{-14}M_{\odot}$, while the nonlinearity 
has been found numerically to arise  
only for the axion stars with masses larger than $\sim 10^{-9}M_{\odot}$.
We need to 
impose a boundary condition such as 
$h_r(r=0)=0$ for the regularity of the space-time.

Changing the scales such that $\tau=m_at$, $x=m_ar$ and $b=a/m_a$, we rewrite 
the equations as follows,

\begin{eqnarray}
\label{bd}
\ddot{b}&=&\dot{V}\dot{b}+b''+(\frac{2}{x}+V')b'-b\quad,\\
\label{V}
V'&\equiv&\frac{h_t'-h_r'}{2}=\epsilon
(\frac{\int_0^xdxx^2(b'^2+\dot{b}^2+b^2)}{x^2}
-xb^2)
\end{eqnarray}
with $\epsilon=4\pi Gm_a^2$,
where we have expressed $V'$  
in terms of the field $b$, solving eq(\ref{h_t}) and eq(\ref{h_r});
here a dot ( dash ) denote a derivative in $\tau$ ( $x$ ).  
We understand that if the gravitational effect is neglected ( $\epsilon=0$ ),
the equation of $b$ is reduced to a Klein-Gordon equation.
Thus the frequency $\omega$ of the field $b$ receives 
a small gravitational effect of the order of  
$\epsilon$; $\omega=1-o(\epsilon)$.

We look for such a solution \cite{real} that 

\begin{equation}
\label{bB}
b=A_0 B(x)\sin\omega\tau+o(\epsilon)\sin3\omega\tau\quad, 
\end{equation}
where $B(x)$ represents coherent axions 
bounded gravitationally with its spatial extension representing 
the radius of the axion star; $B(x)$ is normalized 
such as $B(x=0)=1$. Later we find that 
$A_0$ is a free parameter determining
a mass or a radius of the axion star.  
The second term is a small correction of the order of
$\epsilon$. Here we comment that previous solutions \cite{real}
representing axion stars
with larger masses possess 
more oscillating terms such as $\sin{(2n-1)\omega}$ with $n=1,2,,,$ .
Inserting the formula eq(\ref{bB}) into eq(\ref{bd}) and eq(\ref{V}) and 
taking account of the gravitational effects only 
with the order of $\epsilon$, 
we find that    

\begin{equation}
\label{B}
k^2B=B''+(\frac{2}{x}+\epsilon A_0^2(T+\frac{3U'}{4}))B'+
\frac{\epsilon A_0^2(U+v)B}{2}
\end{equation}
with 
\begin{equation}
T\equiv\frac{\int_0^xz^2B^2dz}{x^2}\quad \mbox{and}\quad
U\equiv\int_0^xdy(\frac{\int_0^yz^2B'^2dz}{y^2}-yB^2)\quad,
\end{equation}
where $k^2$ ( $=1-\omega^2$ ) is a binding energy of axions. We have imposed a 
boundary condition for the consistency such that 
$V(x=0)=h_t(x=0)/2=\epsilon A_0^2v\omega \sin2\omega\tau$ ; this is 
the definition of constant $v$ in the above formula.

We can see that the parameter $\epsilon A_0^2$ can take an arbitrary value
and that it represents the gravitational effect of this system.
Namely the mass of the axionic boson star is determined by choosing 
a value of the parameter. Note that the normalization of $B$ has been 
fixed in eq(\ref{B}) although the equation is a linear in $B$.

We may take the value of $v$ without loss of generality such that 
$v=-U(x=\infty)$. Then the inverse $k^{-1}$ of the binding energy 
is turned out to represent
a radius of the axion star; $B$ decays exponentially 
such as $\exp(-kx)$ for $x\to\infty$.  
It turns out from eq(\ref{B}) that 
the choice of small values of $\epsilon A_0^2$ lead to solutions representing 
the axion stars with small masses and large radii $k^{-1}$.

Before solving eq(\ref{B}) numerically, it is interesting to rewrite the 
equation as following. That is, we rewrite the equation by 
taking only dominant terms of a ``potential'', $V_b$, in eq(\ref{B}),

\begin{equation}
\label{Vb}
V_b=\epsilon A_0^2((T+\frac{3U'}{4})B'+\frac{(U+v)B}{2})
\end{equation}
in the limit of 
the large length scale; setting $x=\lambda y$, we take a dominant term  
as $\lambda \to \infty$. 
This corresponds to taking the axion stars 
with their spatial extension being large.

Then, since the dominant term in the limit is 
the last term in eq(\ref{Vb}), 
$U+v\sim \int_x^{\infty}dxxB^2$,
we obtain the following equation,

\begin{equation}
\label{BB}
\bar{B}=\bar{B}''+\frac{2\bar{B}'}{z}+
\frac{\bar{B}\int_z^{\infty}dyy\bar{B}^2}{2}\quad,
\end{equation}
where we have scaled the variables such that 
$B^2=k^4\bar{B}^2/\epsilon A_0^2$ and $x=k^{-1}z$;
a dash denotes a derivative in $z$.
This equation is much simpler than eq(\ref{B}), where we need to find each 
eigenvalue of $k$ for each value of $\epsilon A_0^2$ given, in order to obtain 
solutions of the axion stars with various masses. On the other hand,
we need only to find an appropriate value of 
$\bar{B}(z=0)=\epsilon A_0^2/k^4$ in order to obtain such solutions in 
eq(\ref{BB}).
A relevant solution we need to find is the solution without any nodes.
Obviously, the solution is characterized by 
one free parameter, $k^4$ or $\epsilon A_0^2$, which is related to the mass 
of the axion star. Namely the choice of a value of the mass determines 
uniquely the properties of the axion star,
e.g. radius of the star, distribution of 
axion field $a$, e.t.c..

Although the equation (\ref{BB}) 
governs the axion star only with the large radius, 
we have confirmed by solving numerically original equation (\ref{B}) that 
the stars of our concern can be controlled 
by the equation (\ref{BB}).

%It means that our solutions represent
%the axion stars with small masses, e.g. $10^{-12}M_{\odot}$; 
%their gravitational fields are sufficiently weak and 
%amplitudes $a$ are much small. Thus nonlinearity of the axion potential
%is irrelevant; we have found that the nonlinearity arises  
%only for the axion stars with masses larger than $\sim 10^{-9}M_{\odot}$.
%The field $a$ of our solutions possesses only one oscillation mode 
%with its frequency approximately given by $m$.
%Other oscillation modes, 
%which exist in general solutions representing the axion stars 
%with larger masses, 
%appear gradually as the masses increase. 

We have confirmed that our numerical solutions may be approximated by
the explicit formula,

\begin{equation}
\label{a}
a=f_{PQ}a_0\sin(m_at)\exp(-r/R_a)\quad, 
\end{equation}
where $t$ ( $r$ ) is time ( radial ) coordinate and 
$f_{PQ}$ is the decay constant of the axion. 
The value of $f_{PQ}$ is constrained from cosmological 
and astrophysical considerations\cite{text,kim} such as 
$10^{10}$GeV $< f_{PQ} <$ $10^{12}$GeV. Corresponding to this constraint,
$m_a$ is constrained roughly such as $10^{-5}$eV $<m_a<$ $10^{-3}$eV.

In the limit of the small mass of the axion star we 
have found a simple relation \cite{iwaza} between the mass, $M_a$ 
and the radius, $R_a$ 
of the axion star, 

\begin{equation}
\label{mass}
M_a=6.4\,\frac{m_{pl}^2}{m_a^2R_a}\quad,
\end{equation} 
with Planck mass $m_{pl}$.
Numerically, for example,  
$R_a=1.6\times10^{10}m_5^{-2}M_{14}^{-1}\mbox{cm}$; hereafter 
we use the notation,  
$M_{n}\equiv M_a/10^{-n}M_{\odot}$ 
and $m_{n}\equiv m_a/10^{-n}\mbox{eV}$. 
A similar formula has been obtained in the case of boson stars of complex 
scalar fields\cite{re}.
We have also found an explicit relation \cite{iwaza} 
between the radius and the dimensionless amplitude $a_0$ in eq(\ref{a}),

\begin{equation}
\label{a_0}
a_0=1.73\times 10^{-8} \frac{(10^8\mbox{cm})^2}{R_a^2}\,
\frac{10^{-5}\mbox{eV}}{m_a}\quad.
\end{equation}
These explicit formulae are used 
for the evaluation
of the dissipation energy of the axion stars in the magnetized 
conducting media.

\section{Axion Star in Magnetic Field}

We now proceed to explain that the axionic boson stars
generate an electric field
in an external magnetic field. It will turn out below that the field 
gives rise intriguing astrophysical phenomena. Thus 
it is important to understand the mechanism of producing the field.
The point is that the axion couples\cite{kim} with the electromagnetic fields
in the following way,

\begin{equation}
   L_{a\gamma\gamma}=c\alpha a\vec{E}\cdot\vec{B}/f_{PQ}\pi
\label{EB}
\end{equation}
with $\alpha=1/137$, where 
$\vec{E}$ and $\vec{B}$ are electric and magnetic fields respectively. 
The value of $c$ depends on the axion models\cite{DFSZ,hadron};
typically it is the order of one.

It follows from this interaction that Gauss law is given by  

\begin{equation}
\label{Gauss}
\vec{\partial}\vec{E}=-c\alpha \vec{\partial}\cdot(a\vec{B})/f_{PQ}\pi
+\mbox{``matter''}
\end{equation}
where the last term ``matter'' denotes contributions from ordinary matters.
The first term in the right hand side 
represents a contribution from the axion.
Thus it turns out that
the axion
field has an electric charge density, 
$\rho_a=-c\alpha\vec{\partial}\cdot(a\vec{B})/f_{PQ}\pi$, 
under the magnetic field $\vec{B}$\cite{Si}.
This charge density does not vanish only when the field configuration $a$
is not spatial uniform 
since $\vec{\partial}\vec{B}=0$.  
Thus the electric field, $E_a$ associated with this axion charge
is produced such that 
$\vec{E_a}=-c\alpha a\vec{B}/f_{PQ}\pi$.  
Note that both of $\rho_a$ and $E_a$ oscillate with the frequency given by 
the mass of the axion in the case of the axion star since 
the field $a$ itself oscillates.

Obviously,
this field induces an oscillating electric current $J_m=\sigma E_a$
in magnetized conducting media with electric conductivity $\sigma$. 
In addition to the current $J_m$ carried 
by ordinary matters, e.g. electrons, there appears 
an electric current, $J_a$, associated with the oscillating charge $\rho_a$
owing to the current conservation\cite{Si} 
( $\partial_0\rho_a-\vec{\partial}\vec{J_a}=0$ ). 
This is given such that 
$\vec{J_a}=-c\alpha\partial_{t}a\vec{B}/f_{PQ}\pi$. 
It is important to note that 
this electric current is present even in nonconducting media like 
vacuum as far as the axion star is exposed to the magnetic field.
On the other hand, current $J_m$ is present 
only in the magnetized conducting media.

Since $\partial_t a\sim m_aa$ in the case of the axion star, 
the ratio of $J_m/J_a$ is given by $\sigma/m_a$. Hence, 
$J_a$ is dominant in the media with $\sigma < 10^{12}/s$, while $J_m$
is dominant in the media with $\sigma > 10^{12}/s$; 
note that $10^{10}/s < m_a < 10^{12}/s$ 
corresponding to the above constraint on $f_{PQ}$.
Astrophysically, insides of neutron stars or white dwarfs\cite{star}, 
which are our concerns in this paper, possess electric conductivities 
large enough for 
$J_m$ to be dominant. On the other hand, envelopes or surfaces of the 
white dwarfs may have the small conductivities so as for $J_a$ to be dominant, 
although the envelopes of the neutron stars 
have still much large conductivities so that $J_m$ is dominant.

Now we discuss some implications of these electric currents.
First we show that these electric currents, especially, $J_m$ yield 
thermal energies to the conducting media owing to Joule's heats.
It implies that the axion stars dissipates their energies in the 
media. It also implies that the media increase their temperatures and 
as a result they radiate thermal photons more than before. In particular,
it will be shown in next section that 
the dark white dwarfs become bright to be detectable,  
when they collide with the axion stars. Such white dwarfs have been supposed 
to be plausible candidates of MACHO and to be cooled sufficiently
so that the optical detection of them are difficult without the heating 
by the collision.

Second we show that since the currents are oscillating, radiations 
are emitted. However, 
most of the radiations are absorbed inside of 
the conducting media. Thus observable radiations are only ones emitted 
around the surfaces of the media. In particular, 
the emission from the surfaces of the neutron
stars and the white dwarfs is important to be discussed.
But there exists a problem that we can not estimate precisely its luminosity 
since we do not have enough information 
about physical properties of the surfaces, e.g. electric conductivity,
opacity, e.t.c. of the magnetized stars. Hence our estimation of 
the luminosity is necessarily ambiguous. 
However, it will be turned out that sufficient amount of the emission for 
observation is expected in the case of the neutron stars since  
electric conductivities of the surfaces ( so called envelopes ) 
are much larger than those of normal metals.

In order to see these expectations,   
first of all, we would like to calculate 
the dissipation energy of the axion star
exposed to an magnetic field. The field is associated with the 
neutron star or the white dwarf\cite{star}. We suppose such a situation that 
these magnetic stars collide with an axion star. As the radius of the 
axion star depends on $M_a$ and $m_a$, we need to treat two cases separately;
the case of the radius of the axion star being larger than the radius $R$
of the medium, $R_a>R$ and the inverse case, $R_a<R$. Namely,  
 the dissipation arises 
when the media are in inside of the axion star and it arises when the axion 
star is in inside of the media.

Denoting the average electric conductivity of the media 
by $\sigma$ and assuming 
the Ohm law, we find that
the axion star dissipates an energy $W$ per unit time,

\begin{eqnarray}
\label{W}
W_>&=&4\sigma \alpha^2c^2B^2a_0^2R^3/3\pi\quad \mbox{for $R_a>R$}\quad,\\
\label{2w}
&=&5.5c^2\times 10^{31}\mbox{erg/s} 
\,\frac{\sigma}{10^{22}/s}\,\frac{m_a^6}{(10^{-5}\mbox{eV})^6}\,
\frac{M^4}{(10^{-14}M_{\odot})^4}\,\frac{R^3}{(10^9\mbox{cm})^3}
\,\frac{B^2}{(10^6G)^2}\quad,\\
\label{2n}
&=&5.5c^2\times 10^{38}\mbox{erg/s} 
\,\frac{\sigma}{10^{26}/s}\,\frac{m_a^6}{(10^{-5}\mbox{eV})^6}\,
\frac{M^4}{(10^{-14}M_{\odot})^4}\,
\frac{R^3}{(10^6\mbox{cm})^3}
\,\frac{B^2}{(10^{12}G)^2}\quad,\\
W_<&=&\sigma \alpha^2c^2B^2R_a^3a_0^2/8\pi\quad \mbox{for $R_a<R$}\quad,\\
\label{3 w}
&=&4c^2\times 10^{38}\mbox{erg/s}\,\frac{\sigma}{10^{22}/s}\,
\frac{M}{10^{-12}M_{\odot}}\,\frac{B^2}{(10^6G)^2}\quad,\\
\label{3n}
&=&4c^2\times 10^{54}\mbox{erg/s}\,\frac{\sigma}{10^{26}/s}\,
\frac{M}{10^{-12}M_{\odot}}\,\frac{B^2}{(10^{12}G)^2}\quad\,
\end{eqnarray}
with $c\sim 1$.
We have used the formulae eq(\ref{a}), eq(\ref{mass}) 
and eq(\ref{a_0}). 
Since the field $a$ 
oscillates\cite{iwaza,real} with a frequency given by
the mass of the axion $m_a$, we have 
taken an average in time over the period, $m_a^{-1}$. 
When $R$ is much smaller than
$R_a$, we have set, $\exp(-r/R_a)=1$ in eq(\ref{a}) in the derivation of 
the formula eq(\ref{W}); $r<R$.
$W$ in the equations eq(\ref{2w}) and eq(\ref{3 w}) is 
for the white dwarf\cite{star} with 
$R\sim 10^9$cm and $B\sim 10^6$G, and 
$W$ in the equations eq(\ref{2n}) and eq(\ref{3n}) is 
for the neutron star\cite{star} with 
$R\sim 10^6$cm and $B\sim 10^{12}$G, respectively. 
The values of $\sigma$ have been 
taken tentatively.

Note that $R_a=1.6\times 10^{10}m_5^{-2}M_{14}^{-1}$cm. 
Hence formula $W_<$ is applied to the white dwarf with $R=10^9$cm
only when $m_5^2 M_{14}>10^2$, e.g. 
$m_5>10$ and $M_{14}>1$. On the other hand,
$W_<$ is applied to the neutron star with $R=10^6$cm only when 
$m_5^2 M_{14}>10^4$, e.g. $m_5>10$ and $M_{14}>1$

We comment that the formula may be applied to the conducting media where the 
Ohmic law holds even for oscillating electric fields with their frequencies
$m_a= 10^{10}\sim 10^{12}$ Hz. In general the law holds in the media where 
electrons interact sufficiently many times in a period of $m_a^{-1}$ 
with each others or other charged 
particles and
diffuse their energies acquired from the electric field. 
Actually the law  
holds in the white dwarfs and neutron stars of our concerns.

We would like to point out that 
although the electric field 
$\vec{E_a}=-c\alpha a\vec{B}/f_{PQ}\pi$ is much small owing to 
the large factor of $f_{PQ}$,
the amount of the dissipation energy $W$ becomes large
because $W$ is proportional to the volume ( $R^3$ ) of the media or 
the volume ( $R_a^3$ ) of the axion stars.

Next we calculate 
the luminosity of the monochromatic radiations emitted 
around the surface of the magnetized media. They arise 
associated with the oscillation 
of the currents $J_a$ or $J_m$. Here we are only concerned with
the case, $R_a>R$. This is because the radiations are emitted only 
around the surface of the media; it is possible
when the media is in inside of the axion star.

We denote a depth of a region from the surface by $d$, in which 
radiations are emitted and can escape from the magnetized conducting media. 
We also denote an average electric conductivity in the region by $\sigma$.
These values are not well known so that we take them as free parameters.
Noting that only radiations from a semi-sphere facing 
observers can arrive at them, we calculate electromagnetic gauge
potentials $A_i$ of the radiations with an appropriate gauge condition,

\begin{eqnarray}
A_i&=&\frac{1}{R_0}\int_{\mbox{surface}}J_m(t-R_0+\vec{x}\cdot\vec{n})\,d^3x\\
&=&\frac{c\alpha\sigma a_0B_i}{\pi R_0}\int_{\mbox{surface}}
\sin m_a(t-R_0+\vec{x}\cdot\vec{n})\,d^3x\\
&=&\frac{2c\alpha\sigma a_0B_iR}{R_0m_a^2}\,(m_ad\cos m_a(t-R_0)-
2\cos m_a(t-R_0+R-d/2)\sin (m_ad/2))
\end{eqnarray}
where we have integrated it over the region around 
the surface with the depth $d\ll R$. $R_0$ is the distance 
between the observer and the media ( $R_0\gg R$ ).
Here we have used the current $J_m=\sigma E_a$ with the field $a$ in 
the approximate formula eq(\ref{a}) with setting $\exp{(-r/R_a)}=1$;
the media is involved fully in the axion star so that $r/R_a\ll 1$.
On the other hand, the current $J_a$ should be used for $\sigma \le m_a$, 
in which case $\sigma$ should be replaced with $m_a$ in the above formula.
Using the gauge potentials, we evaluate the luminosity of 
the monochromatic radiations with the frequency of $m_a$,

\begin{equation}
\label{surface}
L=\frac{8}{3}(\frac{\sigma}{m_a})^2\,c^2\,a_0^2\,B^2\,R^2\,K^2 \quad,
\end{equation}  
where we have taken an average both in time and the 
direction of the magnetic field.
$K^2$ is given such that 

\begin{eqnarray}
\label{K}
K^2&=&(m_a^2d^2+4\sin^2(m_ad/2)-4m_ad\cos(m_aR)\sin(m_ad/2))/2\\
   &\cong& m_a^2d^2/2 \quad \mbox{for $m_ad \gg 1$}\\
   &\cong& m_a^2d^2(1-\cos(m_aR))\quad \mbox{for $m_ad \ll 1$}\quad.
\end{eqnarray}
In both limit $K^2$ is proportional to $m_a^2d^2$. Thus it turns out that
$L$ is proportional to $\sigma^2d^2$ for $m_a\le \sigma$, or to $m_a^2d^2$ for
$\sigma \le m_a$. 
We should note that the luminosity 
is proportional to the surface area $R^2$ of the magnetized stars; 
it is order of $(10^6)^2\mbox{cm}^2$ for neutron stars or 
$(10^9)^2\mbox{cm}^2$ for white dwarfs.
Thus the quantity is 
enhanced even if a luminosity per unit area in the surface is 
quite weak. This is the point we wish to stress.
As we have stated before, phenomena caused by the axion are 
too faint to be detected 
owing to small factor of $m_a/f_{PQ}$. But in our case we have a large 
factor $R^2m_a^2$ of the order of, for example, 
$10^{18}$ in the case of the white dwarfs.
 
To evaluate numerically the value of $L$ 
we need to know the depth $d$ and the electric conductivity $\sigma$
around the surface. It seems that the quantities depend on 
each physical conditions of the surface of the magnetized stars, 
e.g. temperature, constituents e.t.c.. Thus it is difficult 
to estimate generally the luminosity. In latter sections we discuss it 
by assuming that the depth is given by a penetration depth of the radiations.
Then the depth is written in terms of conductivity $\sigma$.
With this simplification, only the conductivity around the surface 
remains as an ambiguous quantity.

\section{White Dwarf}

White dwarfs\cite{star} are stars in the final stage of their lives 
with intermediate masses, $1\,M_{\odot}\sim 8\,M_{\odot}$. 
As is well known,  
they are composed of C or O with atmosphere of H or He, and 
their mass ( radius ) is typically given by $0.5M_{\odot}$ ( $10^9$cm ).
The pressure in the white dwarfs is dominated by 
pressure of degenerate electrons whose density 
is much larger than normal metals. On the other hand, the internal energy
is dominantly given by kinetic and potential energies of 
nuclei such as C and O. Then,  
radiations from them reduce their 
internal energies stored inside. As a result their temperatures decrease 
and their luminosities become small with time  
because they never generate nuclear energies. Hence old white 
dwarfs are expected to be so dark that it is difficult to observe them.
This fact leads to the natural expectation 
that the white dwarfs in the halo are candidates of
MACHO detected with a gravitational microlense effect. 
Actually some of cooling models\cite{cool} of the white dwarfs 
support this expectation,
although this point is still controversial. There are several arguments 
against this possibility of MACHO being the white dwarf\cite{R}. 

Here we assume that MACHO is just a dark white dwarf with sufficiently 
low temperature whose population is 
$2\times10^{11}M_{\odot}/0.5\,M_{\odot}\sim 4\times10^{11}$; 
note that total mass of the halo is about $4\times 10^{11}M_{\odot}$, 
half of which is expected to be the mass of the white 
dwarfs. We show that the dark white dwarfs become rebright with the collision 
of the axion stars. We also calculate the rate of the collision in the halo.
As a result we
find that the number of the white dwarfs in the halo 
increases discontinuously
with luminosities below a certain luminosity 
( $10^{-5.5}L_{\odot}\sim 10^{-7}L_{\odot}$ )
gained by the white dwarfs with 
the collision.

These white dwarfs may have 
strong magnetic fields typically such as $10^6$G.
Then when they collide with the axion stars, the axion field generates 
currents $J_a$ and $J_m$ in the white dwarfs. Consequently,
thermal energies are produced with the dissipation of the axion field 
energy. The energies ( $\propto\sigma J^2$ )
are expected to be large owing to large 
electric conductivity in the white dwarfs. In particular, inside of 
sufficiently cooled white dwarfs is crystallized 
just like solid metals and reaches a
stage of a fast Debye cooling\cite{star}. 
It means that the old white dwarf has been 
sufficiently cooled\cite{cool} so that 
their core temperature is much lower than $10^4$K,
for example. In such a case,
the conductivity of electrons have been found\cite{con} theoretically to be
large; $\sigma\sim 10^{26}\,(T/10^3$K)/sec at density$\sim 10^7$g/cm$^3$ 
where $T$ represent a core temperature inside of the white dwarf.
    
Then, colliding with the axion star, the white dwarf gains the dissipation 
energy per unit time as follows,

\begin{eqnarray}
W_>&=&5.5c^2\times 10^{35}\mbox{erg/s} 
\,\frac{\sigma}{10^{26}/s}\,\frac{m_a^6}{(10^{-5}\mbox{eV})^6}\,
\frac{M_a^4}{(10^{-14}M_{\odot})^4}\,\frac{R^3}{(10^9\mbox{cm})^3}
\,\frac{B^2}{(10^6G)^2}\quad \mbox{for $R_a > R$ ,}\\
W_<&=&4c^2\times 10^{42}\mbox{erg/s}\,\frac{\sigma}{10^{26}/s}\,
\frac{M}{10^{-12}M_{\odot}}\,
\frac{B^2}{(10^6G)^2}\quad \mbox{for $R_a < R$ ,}
\label{Wb}
\end{eqnarray}
where we have taken the above value of the conductivity.
Noting that a constant $c$ is order of 1, we found that the white dwarf 
gains so much thermal energy when it goes through the axion star.

However, the maximal energy the axion star can deposit per unit time
is the energy possessed by a part of the axion star 
which the white dwarf sweeps per unit time,
when the dwarf is smaller than the axion star ( $R_a > R$ ). 
This is owing to the energy conservation.
As relative velocity, $v$, between the white dwarf and the axion star 
in the halo is 
approximately given by $10^{-3}\times$ light velocity,  
the energy stored in the part can be estimated such as 
$3R^2vM/4R_a^3\sim 10^{35}\mbox{erg/s}\,
(M_a/10^{-14}M_{\odot})^4\,(m_a/10^{-5}\mbox{eV})^6$.
This is smaller than $W_>$ estimated naively.
Therefore, real amount of the energy the white dwarf can gain is at most
given by
$W_{real}\sim10^{35}$ erg/s 
$(M_a/10^{-14}M_{\odot})^4\,(m_a/10^{-5}\mbox{eV})^6$.  
The gain of the energy continues 
until the white dwarf passes through the axion star. 
Thus total energy gained by the white dwarf 
( or dissipated by the axion star ) is 
$2 R_a/v\times W_{real}\sim 10^{38}\mbox{erg}\,
(M_a/10^{-14}M_{\odot})^3\,(m_a/10^{-5}\mbox{eV})^4$ in the case of 
the radius of the axion star being larger than that of the white dwarf. 
This is the energy gained by the white dwarf when it passes the axion star
without trapping the star. If the white dwarf traps the axion star,
the white dwarf may gain larger amount of the energy. Maximally 
all of the mass 
$M_a\simeq 1.8\times 10^{40}\mbox{erg}\,(M_a/10^{-14}M_{\odot})$ 
can be transformed to the thermal energy.

On the other hand, when the dwarf is larger than the axion star,
it is obvious from the formula eq(\ref{Wb})
that the whole energy of the axion star,  
$M_a\simeq 1.8\times 10^{42}\mbox{erg}\,(M_a/10^{-12}M_{\odot})$
is dissipated within one second. Namely such a axion star evaporates 
soon after it enters the white dwarf.

Suppose that the white dwarf with mass $0.5\times M_{\odot}$ and 
radius $10^9$cm is in the stage of the Debye cooling.
Then, their specific heat, $c_v$, per ion
is given approximately by \cite{star} 
$c_v\simeq 16\pi^4\,(T/\theta_D)^3/5$,
where $\theta_D$ is the Debye temperature, typically being $10^7$ K.
Hence the injection of the energy, 
$10^{38}\mbox{erg}\,(M/10^{-14}M_{\odot})^3\,(m_a/10^{-5}\mbox{eV})^4$ 
( or $M_a\simeq 10^{42}\mbox{erg}\,(M_a/10^{-12}M_{\odot})$ ),
increases the core temperature of the white dwarf to 
$\simeq 1.5\times 10^4\mbox{K}\,(M/10^{-14}M_{\odot})^{3/4}\,
(m_a/10^{-5}\mbox{eV})$
( or $\simeq 1.5\times 10^5\mbox{K}\,(M_a/10^{-12}M_{\odot})^{1/4}$ )
when the initial temperature is much less than these ones.
%If surface temperature is the same as 
%this temperature, the luminosity is roughly 
%$10^{-3}L_{\odot}\,(M/10^{-14}M_{\odot})^3\,(m_a/10^{-5}\mbox{eV})^4$.
%This is too naive estimation. 
In order to evaluate the luminosity of the dwarf,
we need to know the surface temperature.
Generally, the surface temperature is 
much lower than the core temperature. It depends on opacity of the 
atmosphere of the white dwarf. Recently it has been shown\cite{hansen} 
that the white 
dwarf with atmosphere of helium has much lower opacity at 
surface temperature ( $<6\times 10^3$ K ) than ones 
estimated previously. Such a white dwarf cools more rapidly.
According to the model\cite{hansen} of the cooling, such white dwarfs of ages 
$\simeq10^{10}$ years
have been cooled with their core temperature $\simeq 2.3\times10^5$ K and 
with corresponding surface temperature $2000$ K; its luminosity is 
$10^{-5.6}L_{\odot}$ which is less than 
the minimum luminosity $\sim 10^{-4.8}L_{\odot}$ of 
a white dwarf observed at present.
Thus white dwarfs with their ages older
than $10^{10}$ years must be cooler than this one. 
It is expected that ages of the halo white dwarf population is around 
$1.2\times 10^{10}$ years.
Hence if the axion star with mass larger than $10^{-12}M_{\odot}$ 
( $10^{-9}M_{\odot}$ ) collides 
with the white dwarf whose radius is larger than that of the axion star,  
then the core temperature reaches more than 
$\simeq 1.5\times 10^5$ K ( $\simeq 8.4\times 10^5$ K ) 
and its luminosity does more than 
$L\simeq10^{-6}L_{\odot}$ ( $\simeq 10^{-4.5}L_{\odot}$ ) 
according to the cooling model; here we have simply extrapolated 
its result to the case of lower luminosity than that 
( $\sim 10^{-5.6}L_{\odot}$ ) addressed in the model. 
It takes about $1.4\times 10^{7}$ years ( $5\times 10^{8}$ years )  
for the dwarf to loose the energy injected.

Now we proceed to estimate the rate of the collision between the white dwarf 
and the axion star in the halo.  
Especially, we are concerned 
with the event rate observed
in a solid angle,
$5^{\circ}\times 5^{\circ}$, for example. We assume that 
as indicated by the recent observations of gravitational
microlensing, the half of the halo
is composed of the white dwarfs with mass 
$M=0.5\times M_{\odot}$ and radius $R=10^9$cm. The other
half is assumed to be composed of the axion stars. 
Total mass of the halo is supposed
to be $\sim 4\times 10^{11}M_{\odot}$. 
Furthermore, the distribution\cite{dis} of the halo 
is taken such that its density 
$\propto (r^2+ 3 R_c^2)/(r^2+R_c^2)^2$ with $R_c=4$ kpc 
where $r$ denotes a radial coordinate with the origin being the center of 
the galaxy ( the final result does not depend practically 
on the value of $R_c=2\sim 8$ kpc ).
Then it is easy to evaluate the event rate of the collisions,

\begin{eqnarray}
\label{rate}
&0.5&\,\mbox{per year}\times\frac{(10^{-14}M_{\odot})^3}{M_a^3}\,
\frac{(10^{-5}\mbox{eV})^4}{m_a^4}\,\frac{\Omega}{5^{\circ}\times5^{\circ}}
\quad \mbox{for}\, R<R_a \\
&5&\times 10^{-5}\mbox{per year}\times\frac{10^{-12}M_{\odot}}{M_a}\,
\frac{\Omega}{5^{\circ}\times5^{\circ}} \quad \mbox{for}\, R>R_a 
\end{eqnarray}
where $\Omega$ is a solid angle.  
We have taken into account the fact that the earth is located at about 
$8$ kpc from the center of our galaxy, simply by 
counting the number 
of the collisions arising in the region
from $r=8$ kpc to $50$ kpc.  
Here we have assumed that the collisions take place with their cross 
section simply given by 
$\pi (10^{10}\mbox{cm})^2\,(10^{-14}M_{\odot}/M_a)^2\,m_5^{-4}$ for 
$R<R_a$ and $\pi (10^9\mbox{cm})^2$ for $R>R_a$, respectively.
Namely the cross section is geometrical one of the axion star
or the white dwarf.
But we expect that actual cross section is much larger than this one; 
a tidal force of the white dwarf may decrease kinetic energy of
the axion star by tearing the star.

We also calculate the rate of the collision in the neighborhood of the 
earth. In particular, we with to see how many the collisions occur
within a volume $(1\mbox{Kpc})^3$ in a year around the earth.
This is because since the luminosity of the white dwarfs expected in the 
collisions is much small, the dwarfs need to be located near the earth
in order to be detected.  
We assume that local density of the halo\cite{text} is given by 
$0.5\times10^{-24}$g $\mbox{cm}^{-3}$, half of which is composed of the white
dwarfs and the other half is composed of the axion stars. 
Then it is easy to find that the rate of the collisions is given by 
$\sim 0.06M_{14}^{-3}m_5^{-4}$ per year for $R<R_a$ and 
$\sim 6\times 10^{-6}M_{12}^{-1}$ per year for $R>R_a$, respectively,
where we used  
the relative velocity being equal to $3\times10^7$cm/sec.
It seems that it is difficult to detect the collisions.
Real population of the white dwarfs and the axion stars in the halo 
may be smallar than one we have assumed in the analysis.
Then the collision rate is smaller than the above value and 
it is more difficult to detect the collision with observation of 
monochromatic radiations discussed below.
However, if we take into account the fact that actual 
collision cross section must be much larger than the geometrical one of 
the axion star due to gravitational attraction, the rate will increase. 
To see it we need to analyze numerically the collision in detail.

Here we with to discuss how many white dwarfs heated with the collision
are present in a region around the earth whose volume is assumed to be 
$(1\mbox{Kpc})^3$. For the purpose we note 
that the white dwarfs heated with the collision 
loose the energies injected, taking many years. Hence even if the collision 
rate is small, the number of such white dwarfs increases with time 
and saturates at a balance point between the decay and the production.
For example it takes $1.4\times 10^7$ years ( $5\times 10^8$ years )
for a white dwarf with 
luminosity $10^{-6}L_{\odot}$ ( $10^{-4.5}L_{\odot}$ )
to loose the energy which is injected by
an axion star with $M=10^{-12}M_{\odot}$ ( $M=10^{-9}M_{\odot}$ ). 
On the other hand, the rate of 
the collision producing the dwarf with the luminosity is 
given by $\simeq 6\times 10^{-6}$ per year 
( $\simeq 6\times 10^{-9}$ per year )
in the region with volume $(1\mbox{Kpc})^3$.
Thus it turns out that there present $84$ ( 3 ) such white dwarfs 
in the region.
The number of such white dwarfs
becomes larger as their luminosities gained with the collision are smaller. 
For example, with the collision of the axion star 
$M=10^{-11}M_{\odot}$, a white dwarf gains 
a luminosity $10^{-5.5}L_{\odot}$ and looses its energy 
in $3.7\times 10^7$ years.
On the other hand, the production rate of the dwarfs is 
$6\times 10^{-7}$ per year. Thus there present approximately $22$ such dwarfs.
Accordingly, we understand that as the mass of the axion stars is smaller,
the energy gained by the white dwarfs is smaller and resultant 
their luminosity is smaller, but the number of such dwarfs is larger.

Although we do not know theoretically the masses of the axion stars, 
observations of dark white dwarfs enable 
us to determine the mass of the axion stars.
Namely if there is a threshold luminosity e.g. 
$10^{-5.5}L_{\odot}$ below which the number of the white dwarfs increases
discontinuously, 
it implies that such white dwarfs with the threshold luminosity are produced 
with the collisions of the axion stars with a corresponding mass
e.g. $10^{-11}M_{\odot}$. In this discussion, we have assumed that old 
white dwarfs dominant in the halo have been cooled sufficiently and 
their core temperature is quite low ( e.g. less than $10^4$ K ).

%Therefore we find that even if the cross section of the axion star itself 
%is adopted as a collision cross section, 
%the rate is large for the collision to
%be detectable although it depends on several unknown parameters. 
%With such a direct collision we will observe 
%a phenomena like a nova, which is a white 
%dwarf revived by the collision.

Until now, we have discussed the thermal radiations caused by the axion star 
which dissipates its energy and makes the white dwarf 
warmer than before. 
 
We proceed to discuss the monochromatic radiations generated 
by the oscillating electric currents. The radiations are ones emitted 
only around the surface of the white dwarf. Otherwise, the radiations 
are absorbed inside of the white dwarf. 
There are difficult problems to estimate amount of such radiations
because the physical parameters ( conductivity, opacity, e.t.c. )
around the surface of 
the cooled white dwarfs have not yet been known so well\cite{cool,hansen}.
Thus our estimation is inevitably ambiguous. However, 
with an assumption that the depth $d$ of the region in which the radiations 
are generated and can go out of the white dwarf, is given by 
a penetration depth of the radiations,
we can obtain 
the amount of the radiations definitely with one ambiguous parameter,
the penetration depth, which is written in terms of the electric conductivity
and the frequency of the radiations. 
 
With use of eqs(\ref{surface}) and (\ref{K}), we find the luminosity 
of the radiations,

\begin{eqnarray}
\label{SL1}
L&\sim&10^{21}\mbox{erg/s}
\,B_6^2\,R_9^2\,\sigma_{20}\,m_5^5\,M_{14}^4\,c^2 \quad 
\mbox{for $m_a\ll \sigma$}\\
&\sim&10^{21}\mbox{erg/s}
\,B_6^2\,R_9^2\,m_5^8\,M_{14}^4\,c^2/\sigma_5^2 \quad 
\mbox{for $m_a\gg \sigma$}
\label{SL2}
\end{eqnarray}
where 

\begin{equation}
B_6=\frac{B}{10^6\mbox{G}},\quad R_9=\frac{R}{10^9\mbox{cm}},\quad
\sigma_{20}=\frac{\sigma}{10^{20}/\mbox{s}},\quad
\sigma_{5}=\frac{\sigma}{10^{5}/\mbox{s}}\quad.
\end{equation}  
Here we have only addressed the case that the radius of the white dwarf is 
smaller than $R_a$.
We have used the penetration depth $d=1/2\pi\sigma$ for $m_a\gg \sigma$, and 
$d=\sqrt{1/2\pi \sigma m_a}$ for $m_a\ll \sigma$, respectively. In both case 
we have simply assumed that 
values of both dielectric constant and magnetic permeability 
are the same as those of the vacuum. Numerically, 
$d\sim 10^{5}$ cm$/\sigma_5$ for $m_a\gg \sigma$ and
$d\sim10^{-5}$ cm $(\sigma_{20}m_5)^{-0.5}$ for $m_a\ll \sigma$. 
We should note that as a source of the radiations, 
matter current $J_m$ carried by electrons is dominant 
for $m_a\ll \sigma$, while axion current $J_a$ is dominant for $m_a\gg \sigma$.
In each case we have represented luminosity  
in eq(\ref{SL1}) and eq(\ref{SL2}), respectively.

These monochromatic radiations are emitted only 
during the collision with the axion star. This is because 
after the white dwarf passes the axion star, there are not any oscillating 
currents around its surface; the currents arises only when the dwarf is
passing the inside of the axion star.
Therefore, the period of the emission continueing is given by
$2R_a/v\sim 3.2\times 10^{10}$cm 
$m_5^{-2}M_{14}^{-1} /(3\times10^7$cm/s) $\sim 10^3\,\mbox{sec}$ 
for $m_5^2M_{14}=1$.

It seems that the luminosity is small for the observation
of these radiations. But it depends heavily on the mass of the axion
especially in the case of $m_a\gg \sigma$. For example if $m_a=10^{-4.5}$eV
where $R\sim R_a$,
then $L\sim 10^{25}$erg/s in the case of $m_a\gg \sigma$, eq(\ref{SL2}).
This may be large enough for the observation. 
Furthermore, we point out that 
there are huge ambiguities in the evaluation of the luminosity
as we have mentioned before. Thus we can not determine definitely whether 
the luminosity of the radiations is large enough for observation, or not.
Although there exist some parameter ranges for the radiations being 
observable, it seems difficult to observe the radiations caused with 
the collision unless the luminosity is so large for the radiation from
the distance $10$ Kpc to be detectable. Note that the rate of the collision
is given approximately by $60$$M_{14}^{-3}m_5^{-4}$ 
per year per $(10\mbox{Kpc})^3$.

In the above discussion we have assumed that the axion star does not receive 
any gravitational effects from the white dwarf in the collision. However, 
actually it receives strong gravitational effects from the white dwarf. 
This is because the mass of the white dwarf is much bigger 
than that of the axion star.
So we need to take into account the effects 
in order to see whether or not some of the assumptions are changed,
in particular, the number of the axion stars which was estimated naively
under an assumption of the stars not decaying 
within the age of the Universe.

First of all, we examine whether or not the axion star decays by 
a tidal force when the 
white dwarf passes near it. We suppose that the axion star decays if 
an energy difference caused by the tidal force between different parts
of the axion star is larger than the binding energy of the axion star.
In particular, we wish to estimate  
how many the axion stars 
survive without decaying in the present Universe whose age is approximately 
$10^{10}$ years.

Suppose that an axion star with mass $10^{-14}M_{\odot}$
is composed of two parts which are bounded 
gravitationally with each other; the distance between the two parts is 
assumed to be $R_a$.
When a white dwarf passes the axion star,
each part receives different gravitational force
owing to the difference of their distances from the white dwarf.
In order to examine whether or not this tidal force tears the axion 
star, we compare an energy difference between an gravitational energy 
received by a part of the axion star from the white dwarf 
and an energy received by 
the other part of the axion star,
with the binding energy $GM_a^2/R_a$ of the axion star. 
It is reasonable to think that 
if the energy difference is larger than 
the binding energy, the tidal force tears the axion star. 
The energy difference is dependent on the impact parameter of 
the collision between the axion star and the white dwarf.
Assuming a relative velocity being $10^{-3}\times$ light velocity,
we find with rough estimation that the axion star decays 
with the tidal force when they 
approach within $10^{15}$cm each other. 
Since the number of the axion stars in the halo is given by 
$\sim 2\times 10^{11}M_{\odot}/10^{-14}M_{\odot}=2\times 10^{25}$, and 
the number of the white dwarfs is 
$\sim 2\times 10^{11}M_{\odot}/0.5\times M_{\odot}=4\times 10^{11}$,
the number of the collisions leading to the decay of the axion stars
is at most the order of $10^{23}$ within the age of the Universe. 
Therefore, it turns out that almost of all axion stars ( more than 
$99$ percent ) have survived
against such collisions. Note that the axion stars with masses larger 
than $10^{-14}M_{\odot}$
can survive even more against the tidal force of the white dwarfs.  
This is because the collision cross section becomes smaller for such 
axion stars and so the decay rate is smaller.

Although we found that there still present quite large numbers 
of the axion stars in our halo, we do not know how actually direct 
collisions between the axion star and the white dwarf take place;
a direct collision implies a collision with
their closest distances being less than $R_a\sim 10^{10}$cm.
Namely its cross section is given by 
$\pi(1.6\times10^{10}\mbox{cm}\,(10^{-14}M_{\odot}/M_a))^2$, i.e.
the geometrical cross section of the axion star itself.
It seems that the gravitational force deforms strongly the configulation
of the axion star, but the coherence of the axion field may holds and 
the generation of the electric currents still arises. Thus we expect 
the heat up of the cooled white dwarfs and 
the resultant emissions of the thermal radiations as well as 
the monochromatic radiations.

%Now we wish to discuss what happens when the axion star is exposed to 
%the dipole magnetic field of the white dwarf located far away 
%from the axion star. The axion star distant from the white dwarf recieves 
%only very weak magnetic field. As it approaches the white dwarf closer,
%the field strength becomes larger. Thus approaching it, the axion star 
%looses its energy by radiating electromagnetic fields, which are 
%caused by the current $J_a$ induced inside of the star.

\section{Neutron Stars}

Now we proceed to discuss the influence on neutron stars of their collision 
with the axion stars.
The neutron star\cite{star} is highly dense nuclear matter composed of 
neutrons and protons, although number of the protons is much less than that of 
neutrons. Since their radius, $R_n$, is typically given by $10^6$cm,
we are only concerned with the case of 
$R_n<R_a\sim 10^8m_5^{-2}M_{12}^{-1}$ cm.
Since there are also highly dense electrons, conductivity 
is quite higher than that of the white dwarf. As is well known,
strong magnetic field is present.
Their strength is typically given by $10^{12}$G.  
Hence, the amount of the energy dissipation of the axion star per unit volume 
in the neutron star is much larger than that in the white dwarf.
But as we mentioned in previous section, the actual amount of 
the energy dissipation is restricted owing to the energy conservation.
It is given by $3R_n^2vM_a/4R_a^3\sim 10^{29}\mbox{erg/s}\,
(M_a/10^{-14}M_{\odot})^4\,(m_a/10^{-5}\mbox{eV})^6$.
Total amount of the energy deposited after the collision
is $10^{32}$ erg $(M_a/10^{-14}M_{\odot})^3m_5^4$.
This is a small fraction of the thermal energy possessed by the neutron star.
Therefore, the collision with the axion star does not affect significantly 
thermal contents in the neutron star, contrary to the case of the white dwarf.

Although the thermal energy does not change so much with the collision, 
detectable amount of radiations from the surface 
of the neutron star arises during the collision with the axion star.
As we mentioned, the axion field of the axion star 
generates an oscillating electric field $E_a$ 
when it is under a magnetic field. Thus 
the oscillating current $J_m=\sigma E_a$ is induced around the surface, 
which generates obviously
the radiations. In the case of the neutron stars the electric conductivity
is so large\cite{con} even at the surface 
that the luminosity of the radiations 
is large enough for them to be detectable.

Using the formula eq(\ref{surface}) and eq(\ref{K}), 
we obtain the luminosity of the radiations,

\begin{equation}
L\sim 10^{27}\mbox{erg/s}
\,B_{12}^2\,R_6^2\,\sigma_{20}\,m_5^5\,M_{14}^4
\end{equation}
with

\begin{equation}
B_{12}=\frac{B}{10^{12}G},\quad
R_6=\frac{R_n}{10^{6}\mbox{cm}},\quad
\sigma_{20}=\frac{\sigma}{10^{20}/\mbox{s}},
\end{equation}
where
we have used the above formulae eq(\ref{mass}) and eq(\ref{a_0}) for 
expressing $a_0$ in terms of the mass $M_a$ of the axion star. 
We have assumed for convenience that the conductivity takes a value such as 
$10^{20}$/s. It is reasonable to take such a value of $\sigma$
because the number density of electrons at the surface must be 
much larger than that of normal metals. Then since 
it is much larger than $m_a$, 
the depth $d$ is taken such as $d=\sqrt{1/2\pi \sigma m_a}$ 
which is the penetration 
depth of the radiations.
This luminosity is that of the monochromatic radiation with the 
frequency, $m_a/2\pi=2.4\times 10^{9}m_5$\,Hz.

The emission continues until the axion star passes through the neutron stars.
It takes $2R_a/v\sim 3.2\times 10^{10}$cm 
$m_5^{-2}M_{14}^{-1} /(3\times10^7$cm/s) $\sim 10^3\,\mbox{sec}$ 
$m_5^{-2}M_{14}^{-1}$,
assuming that the velocity of the axion star is 
$3\times10^7$cm/s, which is the typical 
velocity of matters composing the halo in our galaxy.

The monochromatic radiations emitted at a distance $D$ from the earth
is detected at the earth 
with the following strength,

\begin{equation}
\sim 1\,\mbox{Jy}\,B_{12}^2\,R_6^2\,\sigma_{20}\,m_5^4\,M_{14}^4\,
(D_{kpc})^{-2}\,
\end{equation} 
with $D_{kpc}=D/1\mbox{Kpc}$,
where we have assumed that the frequency $\nu$ of the radiations is 
Doppler broadened with a characteristic width $\Delta \nu \sim 10^{-3}\nu$ 
owing to the velocity of the axion stars. Unit of Jy denotes Jansky;
$1\,\mbox{Jy}=10^{-23}\mbox{erg}\,\mbox{cm}^{-2}\,
\mbox{sec}^{-1}\,\mbox{Hz}^{-1}$.
It turns out that the strength of the radiations is large enough to be 
detectable although the duration $\sim 10^3$sec
of emitting the radiations is very short.
 
In order to estimate how frequent the events of the collisions between 
the neutron star and the axion star occur in a neighborhood of the earth, for 
example, $D\le 1$Kpc, we need to know number density of the neutron stars
including old invisible ones in our galaxy. Assuming the number of the 
axion stars in the halo being given by 
$2\times 10^{11}M_{\odot}/10^{-14}M_{\odot}\sim 10^{25}$ and the 
their uniform distribution within a volume $\sim (50\mbox{Kpc})^3$,
we find that the events in a volume $\sim (1\mbox{Kpc})^3$ 
occur with a rate of the order 
$\sim 10^{-9}$ per year when the number of the neutron stars
in the volume is just one. Here the collision cross section is assumed to be 
given by geometrical one $\pi(10^{10})^2\mbox{cm}^2$ of the axion star. 
Therefore it is impossible in practice to detect such phenomena unless 
the number density of 
the invisible old neutron stars in the neighborhood of the earth is 
much larger than one we expect, or the real collision cross section 
including the gravitational effect of the attraction is much larger than 
the geometrical one.

\section{Discussion}
We have shown that the axionic boson stars dissipate their 
energies in the magnetized conducting media such as 
the white dwarfs or the neutron stars. 
Among them, the old halo white dwarfs with sufficiently low core 
temperature ( $<10^3\mbox{K}\sim 10^4$K ), possible candidates of MACHO,
are heated in this mechanism and emit the thermal radiations. 
Their luminosities have been estimated to achieve 
more than a luminosity $10^{-6}L_{\odot}$ (
$10^{-7.1}L_{\odot}$ )
of a white dwarf when 
$M_a$ is larger than $10^{-12}M_{\odot}$  
( $10^{-14}M_{\odot}$ ); in the case of mass, $10^{-14}M_{\odot}$ 
we have assumed the axion stars  
being trapped to the white dwarf 
and dissipating their whole energy.
These dwarfs loose their energies with the radiations and 
become dark with time, while new such dwarfs are produced with the 
collision of the axion stars.
The masses of almost of all axion stars could be fixed 
to a certain value in a range of
$10^{-14}M_{\odot}\sim 10^{-11}M_{\odot}$ when 
they are produced at QCD phase transition.
Thus the core temperature or the luminosity of the dwarf heated with 
the collision must be almost 
the same as each other. 
Therefore, we expect that there exist a threshold luminosity below which 
the number of the halo white dwarfs increases discontinuously. 
The threshold is 
determined with the mass of the axion stars which collide and heat
the dark dwarfs with sufficiently low temperature. In our estimation
the threshold luminosity is given by 
$10^{-5.5}L_{\odot}\sim 10^{-7}L_{\odot}$ corresponding to the masses of the 
axion stars quoted above.

The threshold luminosity is quite small so that 
the collision need to occur near the earth for the detection of 
such white dwarfs.
Thus we have also estimated the rate of the collision within a volume 
of $(1\mbox{Kpc})^3$ around the earth. 
With assumptions of both populations of the white dwarfs and 
the axion stars being given by half of the halo, 
the rate has been turned out 
to be $\sim 0.06M_{14}^{-3}m_5^{-4}$ per year for $R<R_a$ 
( $6\times 10^{-6}M_{12}^{-1}$ 
per year for $R>R_a$ ), 
when the relative 
velocity between the dwarf and the axion star being equal to 
$3\times 10^7$cm/sec; $R_a=1.6\times10^{10}m_5^{-2}M_{14}^{-1}\mbox{cm}$.
This indicates that the detection of the phenomena like a nova caused with
the collision
would be difficult. 
But in the estimation we have not included the gravitational 
attraction. When we take into account the effect, the collision cross section
will be quite larger than the naive geometrical one 
we have used in the estimation.
So we may expect that the cross section will become large so that 
the actual rate of the collision is large 
for us to be able to observe the 
phenomena. On this point we need to simulate numerically the 
collision and to know how the collision occurs. In particular 
we with to know the collision cross section and also to 
know whether or not the axion star is teared 
by the tidal force of the white dwarf.

We have also shown that monochromatic radiations are emitted 
during the collision between the axion star and 
the magnetized conducting media. The radiations are produced 
around the surface of the media by oscillating current, $J_m=\sigma E_a$. 
Especially, strong radiations are expected 
from neutron stars since their electric conductivity $\sigma$ is still large 
even at their envelope.
However, number density of the neutron stars in our galaxy 
may be much small for 
the rate of the collision to be fairly rare; $\sim 10^{-9}/\mbox{Kpc}^3$ 
per year when their number density is given by $1/\mbox{Kpc}^3$.
Thus it is difficult to observe the monochromatic radiations 
from the neutron stars.

On the other hand, although the collision with white dwarfs does not lead 
to strong radiations such as ones from the neutron stars, 
the number density of the dwarfs is supposed to be 
much larger than that of 
the neutron stars so that the rate of the collision is much larger than one in 
the case of the neutron stars. It is approximately given by 
$0.06/\mbox{Kpc}^3$ per year with use of $M_a=10^{-14}M_{\odot}$ and with 
use of the geometrical cross section ( $=\pi R_a^2$ ).   
Hence unless the luminosity of the radiations is large enough for them 
from the distance $10$ Kpc, for example, to be detectable,
the value is a little bit small 
for the observation. 
Furthermore, our assumption of the population of the white dwarfs and the 
axion stars is dubious. In general the population is 
smaller than one we have assumed. Then the collision rate is 
smaller than one in the above estimation. Hence real rate of the collision 
may be small enough so as for the collision to be undetectable. 
But taking account of gravitational 
attraction between the axion star and the dwarf, real cross section 
may be larger than the naive geometrical one. Thus the rate of the 
collision may be larger than one we have estimated. On this point
we need to analyse numerically the way of the collision in detail.

Finally, we mention 
that our estimation is ambiguous in a sense that 
we do not know many physical parameters associated with the phenomena,
for instance, the precise values of both the mass and the population 
of the axion stars, the mass of the axion itself, 
physical properties of invisible old white dwarfs, 
actual cross section of the collision and how the axion star collides with 
the magnetic stars.
Although there are many ambiguous points in the disscusion,
it is important to note that 
the observation of the monochromatic radiations
makes us determine precisely the mass of the axions and the 
observation of the threshold in luminosity function 
of the halo white dwarfs makes us determine the mass of the axion stars.

Theoretically, if we detect the monochromatic radiations which 
are emitted in a short period,
they are good signal indicating 
that the collision between the axion star and the magnetized media actually
occur. After the detection we expect to see a nova in the direction of 
the radiations; the nova is a white dwarf heated with the energy dissipation
of the axion star in the old white dwarf.

A part of this work have been down when the author has visited 
the Theoretical Physics Group at LBNL. He would 
like to express his thank for useful discussions and comments to 
Professors J. Arafune and B. Hansen, and also for
the hospitality in LBNL as well as in  
Tanashi KEK.

%%%%%%%%%%%%%%%%%%%%%%

\end{document}